\newif\ifuseprd
\newif\ifom
\useprdfalse
\omfalse
\ifuseprd 
\documentclass[preprint,tightenlines,floats,draft,prd,eqsecnum,nobibnotes,
nofootinbib]{revtex4}
\else
\documentclass[12pt,letterpaper,nohyper]{JHEP}
\fi 
\usepackage{amsmath,amssymb}
\usepackage[final]{epsfig}

\allowdisplaybreaks[1]
\ifom
\input dnmacs
\fi 

\def\omt{{\ifom{{\dn\dnhalf 
:}~}\else{{3\!{\footnotesize$\mathbf{{\frown}\llap{\text{\tiny$\prime$}}}$}
{\hbox to -.7ex{\null}\llap{\raise1.3ex\hbox{\tiny{\setbox255=\hbox
{$\mathbf{\smile}$}\copy255\kern-.7\wd255{\raise.5ex\hbox{$\mathbf
{\cdot}$}}}}}}}}\fi}}

\newcommand\skipthis[1]{{}}

\newcommand\ct[1]{{\sf {#1}},}
\newcommand\bt[1]{{\em {#1}},}

\newcommand\phepth[1]{{\tt [\hepth{#1}]}}

\chardef\til=`~

\newcommand\p{\ensuremath{\partial}}
\newcommand\evalat[2]{\ensuremath{\left.{#1}\right|_{#2}}}

\newcommand\lvec[2][]{\ensuremath{\overleftarrow{{#2}_{#1}}}}
\newcommand\rvec[2][]{\ensuremath{\overrightarrow{{#2}_{#1}}}}

\newcommand\apr{{\ensuremath{{\alpha'}}}}

\newcommand\ov{\over}
\newcommand\lam{\lambda}
\newcommand\del{\partial}
\newcommand\OO{{\ensuremath{{\cal O}}}}
\newcommand\sd{{\ensuremath{{\cal D}}}}

\providecommand\putabstract[1]{\ifuseprd\begin{abstract} {#1} 
\end{abstract}%
                           \else \abstract{{#1}} \fi}
\providecommand\plb[3]{{Phys.\ Lett.\ B {\bf {#1}}, {#3} ({#2})}}
\providecommand\npb[3]{{Nucl.\ Phys.\ {\bf B{#1}}, {#3} ({#2})}}
\providecommand\jhep[3]{{J.\ High Energy Phys.\ {\bf #1}({#2}){#3}}}
\providecommand\npps[3]{{Nucl.\ Phys.\ {bf {#1}} Proc.\ Suppl.\ {#3} 
({#2})}}
\providecommand\hepth[1]{{\tt hep-th/{#1}}}
\newenvironment{smaleq}{\ifuseprd\else\small\fi}{}

\ifuseprd
\begin{document} 
\fi 

\title{\LARGE 
$\ast$-Trek II: \\
\Large 
$\ast_n$ Operations, Open Wilson Lines and the Seiberg-Witten Map
}
\ifuseprd
\author{Hong Liu}\email{liu@physics.rutgers.edu}
\affiliation{New High Energy Theory Center \\ 
       Rutgers University \\
       126 Frelinghuysen Road \\
       Piscataway, NJ \ 08854}
\else 
\author{Hong Liu\thanks{\tt liu@physics.rutgers.edu}
\\
New High Energy Theory Center \\
Rutgers University \\
126 Frelinghuysen Road \\
Piscataway, NJ \ 08854 \ USA}
\fi 

\putabstract{

Generalizations of the $\ast$-product (e.g. $n$-ary $\ast_n$ 
operations) appear in various places in the discussion of 
noncommutative gauge theories. These  include 
the one-loop effective action of noncommutative gauge theories, 
the couplings between massless closed and open string modes, and 
the Seiberg-Witten map between the ordinary and noncommutative 
Yang-Mills fields. We propose that the natural way to understand 
the $\ast_n$ operations is through the expansion of an open Wilson line.  
We establish the  connection between an open Wilson line and 
the $\ast_n$  operations and use it to 
(I) write down a gauge invariant effective action for the one-loop
$F^4$ terms in the noncommutative ${\cal N} =4 $ SYM theory;
(II) find the gauge invariant couplings between the noncommutative 
SYM modes and the massless closed string modes in flat space;
(III)  propose a closed form for the Seiberg-Witten map in the $U(1)$
case.
}

\preprint{RUNHETC-00-45\ifuseprd,~\else\\\fi hep-th/0011125}

\ifuseprd
\maketitle
\else
\begin{document}
\fi 

\section{Introduction} \label{sec:intro}

Noncommutative gauge theories  have attracted much interest 
recently. They appear naturally in various  
decoupling limits of the worldvolume theories of D-branes in a background 
NS-NS $B$-field~\cite{cds,dh,ardalan,chuho,scho,sw} and provide 
simplified settings for studying nonlocal effects in string theory.

In noncommutative gauge theories, 
gauge invariance becomes subtle.
When there is only adjoint matter, translations along noncommutative 
directions are a subset of gauge transformations, and thus there are
no gauge invariant local operators in position space~\cite{gross,dasrey}.%
\footnote{Matter in other representations can
sometimes be used to construct gauge invariant local 
operators.\cite{rr,gross2}}
It turns out that noncommutative gauge theories allow a new type of gauge 
invariant objects which are localized in momentum space. They are open 
Wilson lines ~\cite{iikk}.  An open Wilson line 
is gauge invariant~\cite{iikk,ambjorn,dasrey,ghi} 
provided that the distance between the 
end points of the line $\Delta x$ and its momentum $k$  satisfy 
the relation
\begin{equation} \label{gawil}
 \Delta x^{\mu} = \theta^{\mu \nu} k_{\nu}
\end{equation}
where $\theta$ is the noncommutative parameter of the theory (see 
Appendix~\ref{sec:notation} for our notations). 
Attaching local operators which transform 
adjointly under the gauge transformations to an open Wilson line 
also yields gauge invariant operators~\cite{ghi,wadia}.

Recently, we have computed the one-loop four gauge boson scattering amplitude
in noncommutative ${\cal N} = 4$ SYM theory and extracted the corresponding 
contribution to the one-loop effective action in a momentum 
expansion~\cite{lm2}%
\footnote{The calculation in~\cite{lm2} was performed in string theory 
and then reduced to field theory result by taking the $\apr \rightarrow 0$
limit. The computations in field theory 
were later performed in~\cite{zanon} using superspace techniques.}. 
It was found that the one-loop low energy effective 
action involves generalizations of $\ast$-products. To each non-planar 
diagram, depending on the number $n$ of external vertex operator 
insertions on each boundary, there is a corresponding $\ast_n$ $n$-ary 
operation%
\footnote{$\ast_2$, also called $\ast'$ in~\cite{garousi,lm2} was first 
found in~\cite{garousi}. It also appeared in~\cite{sado}.}. 
In particular a careful analysis 
of the one-loop $F^4$ terms found that  the resulting effective action does 
not respect the gauge symmetry of the tree-level action when expressed 
in terms of $\ast_2$ and $\ast_3$ operations. For explicit formulas 
for the $\ast_2$ and $\ast_3$ operations and the definition of 
general $\ast_n$ operations, see Appendix~\ref{sec:exam}.

The $\ast_2, \ast_3$ operations were also found in the 
tree-level closed-open  amplitudes in the presence of a 
$B$-field~\cite{garousi,kiem} and in the Seiberg-Witten  map~\cite{sw} 
between the ordinary and noncommutative gauge field variables~\cite{garousi,
mehen}. $\ast_n$ operations should  also be 
present in the tree-level amplitudes between one closed and $n$ open string 
modes in the presence of a $B$-field.

In this paper we propose that a natural way to understand 
the family of $\ast_n$ operations is  through the 
open Wilson lines%
\footnote{Similar ideas were also explored recently in~\cite{mehen}.}
and study some implications. 
In particular, we introduce a family of gauge invariant operators
by integrating the insertion points of the attached operators 
along the path of a straight open Wilson line.
As an example, if  $\OO_i, i=1, \cdots, n$ are local operators  
which transform adjointly under the gauge transformation, we define
\begin{equation} \label{newg}
\begin{split}
{\cal Q} (k) & =    \int d^4 x  \left( \prod_{i=1}^{n}
 \int^{1}_{0} d \tau_i \right) \; 
 P_{\ast} \left[ W(x,C) \prod_{i=1}^n \OO_i (x + \xi(\tau_i))
 \right] \ast e^{i k \cdot x} \\
& \equiv   \int d^4 x  \,
 L_{\ast} \left[ W(x,C) \prod_{i=1}^n \OO_i (x)
 \right] \ast e^{i k \cdot x} 
\end{split}
\end{equation}
where $W(x,C)$ is a straight open Wilson line~\cite{ghi} with its path 
$C$ parameterized by $\xi(\tau), 0 \leq \tau \leq 1$ and $P_{\ast}$
denotes path ordering with respect to the $\ast$-product.
In the second line we have introduced a short-hand notation $L_{\ast}$
to denote the integrations together with the path ordering procedure.
To the lowest order in the expansion of the Wilson line 
in terms of $\hat{A}$,
it can be shown that
\begin{equation} \label{subop}
{\cal Q} (k)  =    \int d^4 x \ast_{n} \left[\OO_{1} (x), \cdots,
\OO_n (x) \right] \,  e^{i k \cdot x} \, \, + \,\, O(\hat{A}) \ .
\end{equation}

From the relation between~\eqref{newg} and~\eqref{subop} 
we can write down a gauge invariant completion 
of the $F^4$ terms in~\cite{lm2,zanon} using operators of type~\eqref{newg},
the expansion of which in $\hat{A}$ reduces to the results 
of~\cite{lm2,zanon} at the lowest order. Higher order terms 
in the expansion can be explicitly checked by looking at the one-loop
amplitudes of higher number of external vectors  and will be discussed 
in a separate place~\cite{lm3}.

Since the $F^4$ term in ${\cal N} = 4$ noncommutative 
SYM theory also has the interpretation
in string theory as a tree-level process for exchanging the massless 
closed string modes between sources on distant 
branes, the completion helps to identify 
the gauge invariant operators to which the closed string modes couple.
For example, the coupling between the dilaton $\phi$ and the modes of the 
noncommutative ${\cal N}=4$ SYM  is:
\begin{equation} \label{coupdi}
S_{I}  = {1 \ov (2 \pi G_s)}  \int d^4 k \, \sqrt{\det G} \, 
 \phi(-k) \, \OO_{\phi} (k) 
\end{equation}
with 
\begin{equation} \label{odil}
\begin{split}
\OO_{\phi}(k) & =  {1 \ov 4} {\rm Tr} \int d^4 x \, 
L_{\ast} \left[ W(x,C) \hat{F}_{\mu \nu} (x)
\hat{F}^{\mu \nu} (x)
\right] \ast e^{i k \cdot x} \\
& =  {1 \ov 4} {\rm Tr} \int d^4 x \, 
\left[ \hat{F}_{\mu \nu} (x) \ast_2 \hat{F}^{\mu \nu} (x)
+ \theta^{\lam \rho} \del_{\rho} \left(\ast_3 [\hat{F}_{\mu \nu} (x),  
\hat{F}^{\mu \nu} (x), \hat{A}_{\lam} (x)] \right) + \cdots \right] 
\ast e^{i k \cdot x}
\end{split}
\end{equation}
where $G_s$ and $G_{\mu \nu}$ are the open string coupling and metric, 
$\hat{F}$'s are contracted by the open string metric, and we have 
used the $L_{\ast}$ notation defined in~\eqref{newg}. In~\eqref{odil} 
each $\hat{F}$ is integrated separately along the line. The resulting 
operator is different from the one obtained by binding them 
together~\cite{ghi}. The coupling~\eqref{coupdi} is written in 
the momentum space and 
there appears no simple counterpart of it in the position space%
\footnote{That the couplings between noncommutative Yang-Mills  and 
closed string modes ought to be in momentum space rather than in 
coordinate space was pointed out in~\cite{dasrey,ghi}.}.
The above identification should be useful for understanding 
the operator-field correspondence in the noncommutative 
version of the AdS/CFT correspondence~\cite{hi,mr}.

It is tempting to speculate that the $\ast_n$ operations appearing 
in the Seiberg-Witten map may also be attributed to the open Wilson lines. 
At a heuristic level,
this expectation is supported by the following reasoning.    
In noncommutative field theories the elementary quanta are no 
longer point particles; instead they are one dimensional extended 
objects (with no oscillations)
with the property that their extension is proportional to the center 
of mass momentum, i.e. 
given by a relation~\cite{bs,lm},
\begin{equation} \label{dipole}
\Delta x^{\mu} = \theta^{\mu \nu} k_{\nu}
\end{equation}
Intuitively this can be understood as an electric dipole moving in a 
background magnetic field at its lowest Landau level~\cite{bs,sj,yin}.
The coincidence of~\eqref{dipole} with the gauge invariance 
condition~\eqref{gawil} of an open Wilson line 
suggest that we may imagine an elementary quantum macroscopically as 
an open Wilson line of the same momentum.
The  field strength  $F_{\mu \nu}(k)$ of momentum $k$ for an 
ordinary $U(1)$ gauge field is invariant under noncommutative gauge 
transformations and creates an elementary quantum of momentum $k$.
It seems physically appealing that $F^{\mu \nu} (k)$ be 
expressible in terms of an open Wilson line of length~\eqref{dipole} 
with possible operator insertions.

We shall argue that this is indeed the case based on the 
earlier results on the Seiberg-Witten map~\cite{garousi,
cornalba1,ishibashi,okuyama,js,jsw,mehen} and the connection
between the $\ast_n$ operations and the open Wilson lines.
We  propose  a closed form for the expression of an ordinary 
$U(1)$ field strength in terms of the noncommutative Yang-Mills 
field,
\begin{equation} \label{ansatz1}
F_{\mu \nu} (k) = \int d^4 x \, L_{\ast} \left[ \sqrt{\det (1 - \theta 
\hat{F})} 
\left( {1 \ov 1 - \hat{F}  \theta} \hat{F}
\right)_{\mu \nu} (x) \, W(x, C) \right] e^{i k \cdot x} \ .
\end{equation} 
In~\eqref{ansatz1}, $W(x,C)$ is a straight open Wilson line, 
the determinant and rational function of $\hat{F}$  should be understood as
a power series expansion, and $L_{\ast}$ acts on each 
term in the expansion as defined in~\eqref{newg}.  
~\eqref{ansatz1} is gauge invariant by construction and recovers the results
of~\cite{garousi,mehen} upon expanding to cubic order in $\hat{A}$. 
~\eqref{ansatz1} also has the correct small $\theta$ limit, in which case 
the $\ast$-product can be replaced by the Poisson Brackets. However, we are 
unable to prove it to all orders in this paper. 

In the situation where we can ignore the derivatives of the field strength,
equation~\eqref{ansatz1} simplifies and can be written as:
\begin{equation} \label{igd}
F_{\mu \nu} (X(x)) =  \left({1 \ov 1 - \hat{F}  \theta} 
\hat{F}\right)_{\mu \nu} (x) \ 
\end{equation}
where 
\begin{equation}
X^{\mu} (x)  = x^{\mu} + \theta^{\mu \nu} \hat{A}_{\nu} (x) \ .
\end{equation}
When applied to the Born-Infeld action,  \eqref{igd}
gives a direct and  simple derivation of the equivalence between the 
noncommutative and ordinary Born-Infeld actions~\cite{sw}.

The plan of the paper is as  follows. In section~\ref{sec:owlast}, we 
point out  
the close relationship between the  $\ast_n$ operations and a straight 
open Wilson line. In section~\ref{sec:f4}, we propose a gauge 
invariant $F^4$ effective action for ${\cal N}=4$ noncommutative SYM theory
and from it work out the couplings between massless closed 
string modes and the  noncommutative Yang-Mills modes.
In section~\ref{sec:swtr} we motivate a closed form for the Seiberg-Witten 
map in the $U(1)$ case. In section~\ref{sec:bi} we derive the 
Seiberg-Witten map in the 
case of a slowly varying field strength and revisit 
the equivalence between the ordinary and noncommutative Born-Infeld 
actions. In particular we clarify certain aspects of the equivalence. 
We close with some remarks in section~\ref{sec:conc}. We have included 
a number of appendices. In Appendix~\ref{sec:notation} we introduce our
conventions for the $\ast$-product and noncommutative gauge theory. In 
Appendix~\ref{sec:exam} we review the definition for the 
$\ast_n$ operations and give some explicit expressions for $n=2,3$. 
Appendix~\ref{sec:newproof} 
contains an alternative derivation of the Seiberg-Witten map in the 
slowly-varying field case.

\section{Open Wilson Lines and the $\ast_n$ Operations} \label{sec:owlast}

In this section we derive some results about the open Wilson lines
and the $\ast_n$ operations. Similar ideas were also explored  
in~\cite{mehen}.
Our conventions  for the $\ast$-product and  noncommutative gauge 
theories are given in Appendix~\ref{sec:notation}. For a review of the
$\ast_n$ operations refer to Appendix~\ref{sec:exam}.
Throughout
the paper we will use $k_1 \times k_2 \equiv k_{1 \mu} \theta^{\mu \nu}
k_{2 \nu}$. 

A straight open Wilson line of momentum $k$ is given by 
\begin{equation} \label{mwilson}
W_k (C) =   {\rm Tr} \int d^4 x \, W(x,C) \ast e^{i k \cdot x}
\end{equation}
where 
\begin{eqnarray} \label{cwilson}
W(x,C) & = & P_{\ast} \exp \left( i 
\int_0^1 d \sigma \partial_{\sigma} \,
\xi^{\mu} (\sigma) \, \hat{A}_{\mu} (x + \xi (\sigma))
\right) 
\end{eqnarray}
and the path $C$ is a straight line
\begin{equation} \label{path}
\xi^{\mu} (\sigma) =  \theta^{\mu \nu} k_{\nu}  \sigma .
\end{equation}
In the above $P_{\ast}$ denotes the path ordering with respect to
the $\ast$-product.
Under a gauge transformation the Wilson line (\ref{cwilson})
transforms as ($l^{\mu} = \theta^{\mu \nu} k_{\nu}$)
$$
W(x,C) \longrightarrow U(x) \ast W(x,C) \ast U(x + l)^{\dagger} 
$$ 
and the ``momentum-space'' representation (\ref{mwilson}) is 
gauge invariant~\cite{iikk,ambjorn,dasrey,ghi}.
Similarly for any local operator $\OO$ which transforms adjointly under 
the gauge transformation, the following operator is gauge 
invariant~\cite{ghi}
\begin{equation}
\OO(k) =  {\rm Tr} \int d^4 x \; P_{\ast} 
\left[ W(x,C) \OO (y) \right] \ast e^{i k \cdot x} ,
\end{equation}
where $y$ is some point on the path (\ref{path}) of the Wilson line.

We now introduce gauge invariant operators of the type 
\begin{equation} \label{giop}
\begin{split}
{\cal Q} (k) & =    \int d^4 x  \left( \prod_{i=1}^{n}
 \int^{1}_{0} d \tau_i \right) \; 
 P_{\ast} \left[ W(x,C) \prod_{i=1}^n \OO_i (x + \xi(\tau_i))
 \right] \ast e^{i k \cdot x} \\
& \equiv   \int d^4 x  \,
 L_{\ast} \left[ W(x,C) \prod_{i=1}^n \OO_i (x)
 \right] \ast e^{i k \cdot x} 
\end{split}
\end{equation}
where $\OO_i$  transform adjointly under the gauge transformation
and $\xi(\sigma)$ is given by (\ref{path}). In~\eqref{giop} 
we integrated the insertion points of the external operators along 
the path and this may be interpreted as a 
most ``democratic'' way of attaching external operators to an open
Wilson line. When there is only a single operator (i.e. $n=1$
in~\eqref{giop}), the integration can be dropped since insertion at 
any point of the straight line is equivalent to another~\cite{ghi}.
Similarly for $n > 1$, we can choose an arbitrary operator 
and fix its  position at e.g. $\tau=0$.

Equation~\eqref{giop} can be  expanded  in terms of a power series 
in $\hat{A}$,
\begin{equation} \label{expaw}
Q (k) = Q_0 (k) + Q_{1} (k, \hat{A}) + \cdots 
\end{equation}
where $Q_{n} (k)$ involves $n$ powers of $\hat{A}$. For example, 
$Q_0(k)$ is given by
\begin{equation} \label{q0}
Q_0 (k) =  {\rm Tr} \int d^4 x 
\left(\prod_{i=1}^{n} \int^{1}_{0} d \tau_i \right) \; 
P_{\ast} \left[  \prod_{i=1}^n \OO_i (x + \xi(\tau_i))
\right] \ast e^{i k \cdot x}
\end{equation}

Let us now look at the structure of each term in~\eqref{expaw} more 
explicitly. We shall first look at the simpler $U(1)$ case.
Fourier-transforming $\OO_i$ in (\ref{q0}) 
\begin{eqnarray}
\OO_i (x + \xi(\tau_i)) & = &
\int \frac{d^4 k_i}{(2 \pi)^4} \, \OO_i (k_i) e^{-i k_i \cdot (x +
\xi(\tau))} \\
& = & \int \frac{d^4 k_i}{(2 \pi)^4} \, \OO_i (k_i) e^{-i k_i \cdot x
- i (k_i \times k) \tau_i}
\end{eqnarray}
and integrating over $x$, we get 
\begin{equation} \label{q0k}
Q_0 (k) =  \left( \prod_{i=1}^n \int \frac{d^4 k_i}{(2 \pi)^4} 
\right) \,
(2 \pi)^4 \delta^{(4)}(k - \sum_{i=1}^n k_i) \,  
\OO_1 (k_1) \cdots \OO_n (k_n) \,  J_n (k_1, \cdots, k_n) 
\end{equation}
where $J_n$ is given by ($\tau_{ij}= \tau_i - \tau_j$) :
\begin{equation} \label{defjn}
J_n (k_1, \cdots, k_n) = \int_{0}^{1} d \tau_1 \cdots 
\int_{0}^{1} d \tau_n \; \exp \left[ - \frac{i}{2} \sum_{i<j}^n
(k_i \times k_j) (2 \tau_{ij} - \epsilon (\tau_{ij})) \right]
\end{equation}
In reaching (\ref{q0k}) from (\ref{q0}) we have used the momentum
conservation condition $k= \sum_{i=1}^n  k_i$ and the identity 
\begin{equation}
\sum_{i=1}^n (k_i \times k) \tau_i = \sum_{i<j}^n (k_i \times k_j )
(\tau_i - \tau_j) .
\end{equation}
Equation~\eqref{defjn} is 
precisely the momentum space kernel of the 
$\ast_n$ operation on the space of $n$-functions defined in \cite{lm2} (see
Appendix~\ref{sec:exam}).
In coordinate space $Q_0$ (\ref{q0k}) can be written in the form 
\begin{equation} \label{astn}
Q_0 (k) = \int d^4 x \, Q_0 (x) \, e^{ i k \cdot x}, 
\;\;\;\;\;\;\;
Q_0 (x) = \ast_n \left[\OO_{1} (x), \cdots, \OO_{n} (x) \right] \ .
\end{equation}
Using the same manipulations, we find the $m$th order term 
in~\eqref{expaw} can be rewritten as
\begin{smaleq}
\begin{equation} \label{qm}
Q_m (k) = {i^m \ov m!}  \left( \prod_{i=1}^{n+m} 
\int \! \frac{d^4 k_i}{(2 \pi)^4} \right) \,
\OO_1 (k_1) \cdots \OO_n (k_n) M(k_{n+1}) \cdots 
M(k_{n+m}) \,  J_{n + m} (k_1, \cdots, k_{m+n})
\end{equation}
\end{smaleq}
where 
\begin{equation}
M(k_i) = \theta^{\mu \nu} k_{\mu} \hat{A}_{\nu} (k_i)
\end{equation}
and we have omitted the $(2 \pi)^4 \delta^{(4)}(k - \sum_{i=1}^{n+m}
k_i)$ factor inside the integral to make the formula compact. 
In coordinate space, equation~\eqref{qm} can  be written as  
\begin{equation} \label{qmc}
Q_m (x) = {1 \ov m!} (\theta \del)^{\mu_1} \cdots 
(\theta \del)^{\mu_{m}} 
\ast_{m + n} \left[ \OO_{1} (x),  \cdots , \OO_{n} (x), 
\hat{A} _{\mu_1} (x),  \cdots ,\hat{A} _{\mu_m} (x) \right]
\end{equation}
with $(\theta \del)^{\mu} = \theta^{\mu \nu} \del_{\nu}$.

The explicitly expression for $J_n$ and the corresponding $\ast_n$
operation in the $U(1)$ case  
can be found straightforwardly from~\eqref{defjn}. However,
when $n$ becomes large the structure becomes quite complicated and hard
to extract useful  information. Here we list some useful properties:
\begin{enumerate}
\item $\ast_n$ is fully symmetric, i.e. 
\begin{equation}
\ast_{n} (\cdots, f, \cdots, g, \cdots)
=  \ast_{n} (\cdots, g, \cdots, f, \cdots)
\end{equation}
This property also applies to the general $U(N)$ case.
\item $J_n$ satisfies a descent relation,
\begin{equation}
J_n (k_1, \cdots, k_n) = {2 \ov k_n \times k } \sum_{j=1}^{n-1}
\sin {k_n \times k_j \ov 2} J_{n-1} (k_1, \cdots, k_j + k_n, \cdots,
k_{n-1})
\end{equation}
with $k = \sum_{i=1}^n k_i$. 
In position space this implies\footnote{The equations for $n=2,3$ were also
found in~\cite{mehen}.}
\begin{equation}
\theta^{ij} \del_{j} \left[ \ast_n (f_1(x), \cdots, f_n(x), \del_i g(x))
\right] 
= i \sum_{j=1}^{n-1} \ast_{n-1} (f_1(x), \cdots, [f_{j},g], \cdots, 
f_{n-1}(x) ) \ ,
\end{equation}
where 
\begin{equation}
[f, g] = f(x) \ast g(x) - g(x) \ast f(x) \ .
\end{equation}
\end{enumerate}
The explicit expressions  for 
$J_2, \ast_2$ and $J_3, \ast_3$ are listed 
in the Appendix~\ref{sec:exam}.

More generally for a $U(N)$ noncommutative gauge theory, the structures 
of $J_n$ and $\ast_n$ are more complicated, since we have to include 
the ordering of matrices in~\eqref{defjn} and~\eqref{qm}. For example, 
let $\OO_i  = \OO_i^{a_i} T^{a_i}$ where $T^{a_i}$ are a set of basis for 
$N \times N$ Hermitian matrices, then equation~\eqref{defjn} now becomes 
\begin{equation} \label{defnajn}
\begin{split}
& J_n (a_1,k_1; \cdots; a_n, k_n)  \\
& = \left( \prod_{i=1}^n \int_{0}^{1} d \tau_i \right) \,
P_{\tau} \left( T^{a_1} \cdots T^{a_n} \right) \,
\exp \left[ - \frac{i}{2} \sum_{i<j}^n
(k_i \times k_j) (2 \tau_{ij} - \epsilon (\tau_{ij})) \right]
\end{split}
\end{equation}
where $P_{\tau}$ denotes an ordering of matrices according to the 
ordering of $\tau_i$. With the above matrix ordering in mind, the 
formulas~\eqref{astn} and~\eqref{qmc} apply to general $U(N)$ case 
without change. 

Note in $Q_{m}, m=0,1, \cdots$  all the entries (including external
operators and $\hat{A}$'s from the Wilson line) 
are completely symmetric under change of orderings, as a
result of $\ast_n$ operation. Therefore, \eqref{giop} may be
considered  as a ``generalized symmetrized trace'' prescription for 
the algebra of the Moyal product.

\section{One-loop $\hat{F}^4$ Terms
in ${\cal N} =4$ Noncommutative SYM} \label{sec:f4} 
 
\subsection{Gauge Invariant Completion of the One-loop 
$\hat{F}^4$ Terms}

In~\cite{lm2,zanon}, it was found that the non-planar part of 
one-loop $F^4$ terms in ${\cal N} =4$ 
noncommutative Super-Yang-Mills theory involve $\ast_2$ and $\ast_3$ 
operations.  It was also found that if we na\"{\i}vely extend the gauge 
invariant on-shell amplitudes off-shell using the $\ast_2$ and $\ast_3$ 
operations, the resulting effective action does not respect the 
gauge symmetry of the tree-level action. It was speculated there that it 
might be possible to write down a gauge invariant effective action 
using open Wilson lines as they are the natural gauge invariant 
objects of the theory. From the connections between an open Wilson line
and the $\ast_n$ operations found in section~\ref{sec:owlast}, we can now 
immediately write down such a candidate. 
 
Let us first recall the results of~\cite{lm2} (see also~\cite{zanon}). 
Consider a $U(N)$ 
noncommutative ${\cal N}=4$ SYM theory which is broken 
by a Higgs mechanism to $U(N_1) \times U(N_2)$ ($N= N_1 + N_2$). 
The one-loop scattering amplitudes  between massless vector bosons 
in the unbroken subgroups $U(N_1)$ and $U(N_2)$ are given by 
non-planar diagrams with intermediate loop particles  massive 
$W$-bosons (and their super-partners).
With four external vectors the amplitudes at the lowest order in 
momentum expansion (as compared to the mass $m$ of the $W$-bosons) give 
rise to the $F^4$ terms in the low energy  effective action.
The non-planar part of $F^4$ terms found in~\cite{lm2} is:
\begin{subequations} \label{ncf4}
\begin{equation} \label{usencf4}
\begin{split}
\Gamma_{\text{1-loop}} & = -\frac{1}{4! (4\pi)^{2}}
\int d^{4} x \, \sqrt{\det G} \,
t^{\mu\nu\rho\sigma\lambda\tau\alpha\beta} 
\\* & \mspace{-40mu} \left.
\times \biggl\{3{\rm Tr}_{U(N_1)}[F_{\mu\nu}(x) \ast_2 F_{\rho\sigma}(x)]\, 
I_{2}\Bigl(m,\sqrt{\lvec[\mu]{\p}\bigl(\Theta G 
       \Theta\bigr)^{\mu\nu}\rvec[\nu]{\p}}\Bigr)
{\rm Tr}_{U(N_2)}[F_{\lambda\tau}(x) \ast_2 F_{\alpha\beta}(x)] 
\right. \\*
& \mspace{-40mu} - \biggl. 
4 {\rm Tr}_{U(N_1)}
    \bigl(\ast_3[F_{\mu\nu}(x),F_{\rho\sigma}(x),F_{\lambda\tau}(x)]\bigr)
I_{2}\Bigl(m,\sqrt{-\p_\mu \bigl(\Theta G \Theta\bigr)^{\mu\nu}
      \p_\nu} \Bigr)
{\rm Tr}_{U(N_2)}(F_{\alpha\beta}(x))
\biggr\} 
\\* & \mspace{-40mu} + (N_1 \rightarrow N_2)
\end{split}
\end{equation}
where the traces are taken in the fundamental of the indicated 
subgroup and 
\begin{equation} \label{defImp}
I_{2}(m,x) \equiv \left(\frac{x}{2m}\right)^2 K_2(m x) \ .
\end{equation}
\end{subequations}
with  $K_2$ a modified Bessel function. In~\eqref{usencf4}  
$t^{\mu\nu\rho\sigma\lambda\tau\alpha\beta}$ is a tensor composed of 
open string metrics and when contracted with four anti-symmetric tensors
gives the structure 
\begin{equation} \label{ttensor}
\begin{split}
& t^{\mu\nu\rho\sigma\lambda\tau\alpha\beta} 
M_{1 \mu\nu} M_{2 \rho\sigma} M_{3 \lambda\tau}   
M_{4 \alpha\beta}  \\*
& = 2^4 \left( M_{1 \mu}{^\lambda} M_{2 \lambda}{^\alpha} M_{3 \alpha}{^\nu}
M_{4 \nu}{^\mu} - \frac{1}{4} M_{1 \mu \nu } 
M_{2}^{\nu \mu} M_{3 \tau \sigma}
M_{4}^{ \sigma \tau} + \text{2 permutations} \right).
\end{split}
\end{equation}
where the two permutations 
are given by changing the ordering from (1234) to $(1342)$ and 
$(1423)$.

To write down a gauge invariant version of~\eqref{ncf4}, we 
define 
\begin{equation} \label{op1}
\begin{split}
\OO_{\mu \nu \rho \sigma}(k) & =  \int d^4 x \, 
L_{\ast} \left[ W(x,C) \hat{F}_{\mu \nu} (x)
\hat{F}_{\rho \sigma} (x)
\right] \ast e^{i k \cdot x} \\
\OO_{\mu \nu \rho \sigma \lam \tau}(k)& =  \int d^4 x \, 
L_{\ast} \left[ W(x,C) \hat{F}_{\mu \nu} (x)
\hat{F}_{\rho \sigma} (x) \hat{F}_{\lam \tau} (x)
\right] \ast e^{i k \cdot x} \ ,
\end{split}
\end{equation}
where according to~\eqref{newg} each $\hat{F}$ is integrated 
separately along the line%
\footnote{Note in ${\cal N}=4$ noncommutative SYM theory, the open 
Wilson lines in~\eqref{op1} might also contain 
scalar fields $\Phi$ or fermions in addition to~\eqref{cwilson}, 
e.g.~\cite{malda,yee,dasrey}
\begin{equation}
W (x,C) = P_{\ast}  \, \exp \left[i 
\int_0^1 d \sigma \left( \partial_{\sigma} \,
\xi^{\mu} (\sigma) \, \hat{A}_{\mu} (x + \xi (\sigma))
+ |\partial_{\sigma} \,\xi| \, \Omega \cdot \Phi(x + \xi (\sigma))
\right) \right] 
\end{equation}
with $\Omega$ a unit vector on $S^5$.}.
From equation~\eqref{q0k} and~\eqref{astn}, the $\hat{F}^4$ effective 
action~\eqref{ncf4}
can be recovered from the lowest order term of 
the following gauge invariant action 
\begin{equation} \label{f4an}
\begin{split}
\Gamma_{{\rm 1-loop}} & 
= -{1 \ov 4 ! (4 \pi)^2} \int \frac{d^4 k}{(2 \pi)^4} \, 
t^{\mu \nu \rho \sigma \lambda
 \tau \alpha \beta} \, \big[ 3 \, {\rm Tr}_{U(N_1)} 
\OO_{\mu \nu \rho \sigma} (k)
 \,  I_{2} (m, |\Delta x|) \,  {\rm Tr}_{U(N_2)} 
\OO_{\lambda \tau \alpha \beta} (- k) \\
& -  4 \, {\rm Tr}_{U(N_1)} \OO_{\mu \nu \rho \sigma \lambda \tau} (k) \, 
I_{2} (m, |\Delta x| ) \, {\rm Tr}_{U(N_2)} 
\hat{F}_{\alpha \beta} (-k) \big] \,\,  + \,\,  (N_1 \leftrightarrow N_2)
\end{split} 
\end{equation}
where $ |\Delta x| = |\Theta_{\mu \nu} k_{\nu}|$ and 
$I_2 (m, |\Delta x|)$ is given by~\eqref{defImp}.

In the above we have written the gauge invariant effective action in 
momentum space as the gauge invariant operators~\eqref{op1} 
are well-defined local operators in momentum space. They 
do not appear to have sensible Fourier transformations to position  space. 
This is a reflection of the fact that translations are part of the gauge 
symmetry~\cite{dasrey,gross}. 
The higher order terms in~\eqref{f4an} can be checked 
explicitly by computing the higher-point amplitudes
and will be discussed elsewhere~\cite{lm3}.  

\subsection{Closed String Couplings to the Noncommutative Yang-Mills 
Modes} 

The effective action~\eqref{ncf4}, ~\eqref{f4an} can 
also be understood as a tree-level process for exchanging 
closed string modes between  gauge invariant sources on distant branes%
\footnote{For discussions in ordinary flat space or in the context 
of AdS/CFT correspondence, see e.g.~\cite{douglas,das,bak}.}.
It is well known that the contributions from massive modes to $F^4$ terms 
cancel among themselves~\cite{DKPS} and only the massless modes are exchanged.
From this point of view, $I_{2} (m, |\Delta x|)$ is nothing but 
the propagator for a masless scalar field in the transverse dimensions.
The intermediate closed string modes 
have momentum $k_{\mu}$ along the brane directions, which appears as an
effective mass term for the transverse propagator,
\begin{equation} \label{transprop}
G(k, r) = {1 \ov \sqrt{\det g}} \int \frac{d^{\tilde{d}} q}
{(2 \pi)^{\tilde{d}}}
\frac{e^{i q \cdot r}}{q^2 + M^2}.
\end{equation}
\begin{equation} \label{mcl}
M^2 = -k_\mu g^{\mu\nu} k_\nu = -k_\mu G^{\mu\nu} k_\nu +
\left(\frac{1}{2\pi\apr}\right)^2 k_\mu (\theta G \theta)^{\mu\nu} k_\nu 
\sim \left(\frac{\Delta x}{2\pi\apr}\right)^2 \ .
\end{equation}
In~\eqref{transprop}, $\tilde{d}$ is the dimension of the transverse space,
the distance between the two stacks of D-barnes is $r = 2 \pi \apr m$ 
and we have used the relation between the closed ($g_{\mu \nu}$) 
and open string metrics ($G_{\mu \nu}$)~\cite{sw},
\begin{equation} \label{clop}
g^{\mu \nu} = G^{\mu \nu} - \left( {1 \ov 2 \pi \apr} \right)^2 
(\theta G \theta)^{\mu \nu}
\end{equation}
Note that we are considering the field theory limit in which $m$ and 
$G, \theta$ are finite while $\apr, r \rightarrow 0$.
It is easy to check from~\eqref{transprop} and~\eqref{defImp} that 
\begin{equation} \label{propeq}
I_2 (m, |\Delta x|) = \sqrt{\det g} \, 2 \pi^3 \, ( 2 \pi \apr)^4 \,  
G(k, r) \ .
\end{equation}

Using the explicit tensor structure of $t$ in~\eqref{ttensor}, the effective 
action~\eqref{f4an} can be rewritten as
\begin{equation} \label{sugraac}
\begin{split}
\Gamma_{{\rm 1-loop}} = -2 \kappa_{10}^2  \det G {1 \ov (2 \pi G_s)^2}  
& \int d^4 k \, G(k, r) \,  \biggl[ \OO^{(1)}_{\phi} (-k) \,
 \OO^{(2)}_{\phi} (k)
+ \OO^{(1)}_{\chi} (-k) \,  \OO^{(2)}_{\chi} (k) \\
& + {1 \ov 2} T^{(1) \mu}_{\nu} (-k) \, T_{\mu}^{(2) \nu} (k)
+ {1 \ov 2}  \Sigma^{(1) \mu}_{\nu} (-k) \,
\Sigma_{\mu}^{(2)\nu} (k)|_{4} \biggr] \ .
\end{split}
\end{equation}
where 
$$
2 \kappa_{10}^2 = (2 \pi)^7 g_s^2 \apr^4 \ , 
$$
and $g_s$ and $G_s$ are closed and open string couplings, which are 
related by~\cite{sw}
$$
G_s = g_s \left( {\det G \ov \det g} \right)^{1 \ov 4} \ .
$$
The various gauge invariant operators in~\eqref{sugraac} are defined 
by
\begin{eqnarray} \label{dila}
\OO_{\phi}(k) & = &  {1 \ov 4} {\rm Tr} \int d^4 x \, 
L_{\ast} \left[ W(x,C) \hat{F}_{\mu \nu} (x)
\hat{F}^{\mu \nu} (x)
\right] \ast e^{i k \cdot x} \\
\OO_{\chi}(k)& = & {1 \ov 8} \epsilon_{\mu \nu \lam \rho} 
{\rm Tr}  \int d^4 x \, L_{\ast} \left[ W(x,C) \hat{F}^{\mu \nu} (x)
\hat{F}^{\lam \rho} (x)
\right] \ast e^{i k \cdot x} \\
T^{\mu}_{\nu} (k) & = &  {\rm Tr} 
\int d^4 x \, L_{\ast} \left[ W(x,C) \left(\hat{F}^{\mu}_{\alpha} (x)
\hat{F}_{\nu}^{\alpha} (x) - {1 \ov 4} \delta^{\mu}_{\nu} \hat{F}^{\lam \rho}
\hat{F}_{\lam \rho} \right)
\right] \ast e^{i k \cdot x} \\
\label{ant}
\Sigma^{\mu}_{\nu} (k) & = &  {\rm Tr} 
\int d^4 x \, L_{\ast} \left[ W(x,C) \left(\hat{F}^{\mu}_{\nu} + 
\hat{F}^{\mu}_{\alpha} 
\hat{F}_{\beta}^{\alpha} \hat{F}^{\beta}_{\nu}  
- {1 \ov 4} \hat{F}^{\lam \rho} \hat{F}_{\lam \rho}
\hat{F}^{\mu}_{\nu} \right)
\right] \ast e^{i k \cdot x}
\end{eqnarray}
Recall that $L_{\ast}$ was defined in~\eqref{newg} and that in the 
non-Abelian case it also includes the matrix orderings.
In the above definitions, the indices are raised and lowered by the open 
string metric. In~\eqref{sugraac}, the 
superscripts $(i), i=1,2$  corresponds to taking trace 
in $U(N_i), i=1,2$ respectively and  the subscript $|_4$ in the last 
term  of~\eqref{sugraac} means taking only the terms with four factors 
of the gauge field. The expansions of the 
operators~\eqref{dila}--\eqref{ant} in $\hat{A}$ can be found 
using~\eqref{qm} and~\eqref{qmc}.

Equation~\eqref{sugraac} can be obtained from tree-level exchanges of 
the following momentum space action describing the couplings between 
the open and the closed string modes
\begin{equation} \label{coupling}
\begin{split}
S_{I}  = {\sqrt{2} \kappa_{10} \ov (2 \pi G_s)}  \int d^4 k \, 
\sqrt{\det G} \, & 
\bigl[ \phi(-k) \, \OO_{\phi} (k) + \chi (- k) \, 
\OO_{\chi} (k) \\
& + {1 \ov 2} h_{\mu}^{\nu} (-k) \, T^{\mu}_{\nu} (k) 
+ {1 \ov 2} b^{\mu}_{\nu} (-k) \, \tilde{\Sigma}_{\mu}^{\nu} (k) \bigr]
\end{split}
\end{equation}
where $\phi$, $\chi$, $h^{\mu}_{\nu}$, $b_{\mu}^{\nu}$  
are small fluctuations of the dilaton, axion, graviton (polarized along 
the brane directions) and NS-NS $2$-form  respectively. 
In~\eqref{coupling} the closed string modes are canonically normalized 
and $\tilde{\Sigma}_{\mu}^{\nu}$ is given from~\eqref{ant} by
\begin{smaleq}
\begin{equation}
\tilde{\Sigma}^{\mu}_{\nu} (k)  =   {\rm Tr} 
\int d^4 x \, L_{\ast} \left[ W(x,C) \left(
({1 \ov 2 \kappa_{10}^2})^{1 \ov 4}\hat{F}^{\mu}_{\nu} +  
(2 \kappa_{10}^2)^{1 \ov 4} (
\hat{F}^{\mu}_{\alpha} 
\hat{F}_{\beta}^{\alpha} \hat{F}^{\beta}_{\nu}  
- {1 \ov 4} \hat{F}^{\lam \rho} \hat{F}_{\lam \rho}
\hat{F}^{\mu}_{\nu}) \right)
\right] \ast e^{i k \cdot x} \ .
\end{equation}
\end{smaleq}
Note while the 
indices of $T^{\mu}_{\nu} 
$ and $\Sigma_{\mu}^{\nu}$ are raised and lowered by the open string 
metric $G_{\mu \nu}$ , those of $h_{\mu}^{\nu}$ and $b^{\mu}_{\nu}$ are 
raised and lowered by the closed string metric $g_{\mu \nu}$. So we should 
be careful that for instance the coupling $h_{\mu}^{\nu} T^{\mu}_{\nu}$ is
different from  $h_{\mu \nu} T^{\mu \nu}$.

\eqref{coupling} gives the couplings of the lowest dimension gauge 
invariant operators in ${\cal N} = 4$ noncommutative SYM 
to massless closed string modes in a flat space background.
This identification should be helpful for understanding the supergravity 
description of the ${\cal N} =4$ noncommutative SYM theory~\cite{hi,mr}. 
We note that the operators~\eqref{dila}--\eqref{ant} may 
also have dependence on  the scalar fields and fermions of 
the ${\cal N} =4$ theory. It would be interesting to work them out by 
supersymmetry  or by looking at the amplitudes involving scalars and 
fermions.

\section{The Seiberg-Witten Map} \label{sec:swtr}

In this section we propose a closed form for the Seiberg-Witten 
map~\cite{sw} between  the ordinary and noncommutative Yang-Mills 
field. We shall 
only discuss the simpler $U(1)$ case. Our proposal is based on the 
earlier works on the Seiberg-Witten map~\cite{garousi,cornalba1,
ishibashi,okuyama,js,jsw,mehen} and the connection between open Wilson 
lines and $\ast_n$ operations discussed in section~\ref{sec:owlast}.
We shall first derive the map to first order in $\theta$, but to all 
orders in $\hat{A}$, using the results of~\cite{cornalba1,ishibashi,js}, 
and then try to generalize it to all orders in $\theta$.
For definiteness we will restrict our discussion to four Euclidean dimensions,
although our results in this and next sections apply 
invariably to any dimensions.

\subsection{The Seiberg-Witten Map to First Order in $\theta$} 
\label{sec:poisson}

Mathematically speaking, the Seiberg-Witten map can be understood as 
the transformation between star-products associated with  
cohomologically equivalent symplectic forms~\cite{js,jsw}. 
The first part of this subsection is a review of~\cite{ishibashi,cornalba1,
cornalba2,js}. Consider two symplectic forms 
\begin{equation} \label{defsm}
 \omega_{\mu \nu}  = (\theta^{-1})_{\mu \nu} + F_{\mu 
 \nu},\;\;\;\;\;\;
 B_{\mu \nu}  = (\theta^{-1})_{\mu \nu}
\end{equation}
 where $\theta^{\mu \nu}$ is a constant anti-symmetric tensor 
 and $F = dA$ is the field strength of a Abelian gauge field $A_{\mu}$. 
 We shall assume that both $\omega$ and $B$ are non-degenerate.
 We can associate  star products $\ast_{\omega}$ and $\ast_{B}$
 with $\omega$ and $B$ respectively. $\ast_B$ is  the standard Moyal 
 product with a noncommutative parameter $\theta^{\mu \nu}$, while
 for general $F$, $\ast_{\omega}$ must be defined a la 
 Kontsevich~\cite{kont} and is much more complicated.
 
 Since $B$ and $\omega$ differ by an exact form, it is possible to find 
 a coordinate transformation $\lam$ which maps $\omega$ to $B$, i.e. 
 $\lam: x \rightarrow y = y(x)$ so that 
 \begin{equation} \label{diffeo} 
 {\del y^{\rho} \over \del x^{\mu}}
 {\del y^{\lam} \over \del x^{\nu}} \omega_{\rho \lam} (y) = B_{\mu \nu} (x) 
 ={\rm constant} \ .
 \end{equation}
 Thus the symplectic structures defined by $\omega$ and $B$ belong to the 
 same  equivalence class  and the two star products $\ast_{\omega}$ and 
 $\ast_{B}$ must also be equivalent. More explicitly, there exists 
 a map $\sd$ acting on the space of functions which satisfies
 \begin{equation} \label{defD}
 \sd(f \ast_{\omega} g) = \sd f \ast_{B} \sd g \ .
 \end{equation}
 In particular the noncommutative Yang-Mills field is defined by
 \begin{equation} \label{defA}
 X^{\mu} (x) = \sd y^{\mu} \equiv x^{\mu} + \theta^{\mu \nu} \hat{A}_{\nu} \ .
 \end{equation}
It can be shown that an ordinary gauge transformation of $A$ induces 
a noncommutative gauge transformation of $\hat{A}$ and vice  
versa~\cite{js}. Heuristically, the physical information originally 
contained in $F$ is transferred to  $\hat{A}$ through  the map $\sd$.

 Although in principle  $\sd$ can be worked out in the formalism of 
 Kontsevich~\cite{kont} order by order in $\theta$ 
 and $A$ (see e.g.~\cite{js,jsw}), its structure appears to be quite
 complicated. In this paper we pursue a slightly different route. We first 
 find the map at first order in $\theta$ but to all orders in $\hat{A}$ and
 then try to generalize it to all orders in $\theta$. 
  
 To first order in $\theta$, $\sd$ is given by $\sd = \lam^*$~\cite{ishibashi,
 cornalba1,js}, where $\lam$ is the  diffeomorphism~\eqref{diffeo}
 which maps $\omega$ to $B$. At this order,~\eqref{defD} reduces to an 
 equivalence between the  associated Poisson brackets
 \begin{equation} \label{pore}
 \lam^{*} \{f, g \}_{w} = \{\lam^* f, \lam^* g \}_{B},
 \end{equation}
 where $\lam^* f = f \circ \lam = f (y(x))$ and $\{\}_{\omega}$, 
 $\{\}_{B}$ are the Poisson Brackets 
 with respect to $\omega$ and $B$ respectively. 
 The counterpart of~\eqref{defA} is 
 \begin{equation} 
 y^{\mu} (x) = \lam^* y^{\mu} = x^{\mu} + \theta^{\mu \nu} a_{\nu}(x) .
 \end{equation}
 where we have used a different symbol $a_{\mu}$ to denote the noncommuative 
 gauge field $\hat{A}_{\mu}$ at the lowest order in $\theta$. 
 
In~\eqref{pore} setting $f(y) = y^{\mu}, \, g(y) = y^{\nu}$, we obtain that  
 \begin{eqnarray} \label{deff}
 (\omega^{-1})^{\mu \nu} (y (x)) & = & \{x^{\mu} + \theta^{\mu \sigma}
 a_{\sigma}(x), x^{\nu} + \theta^{\nu \rho} a_{\rho}(x) \}_{B} =
 - \left( \theta (f - \theta^{-1} ) \theta \right)^{\mu \nu} \ ,
 \end{eqnarray}
 with the corresponding field strength for $a$ given by 
 \begin{equation} \label{pofs}
 f_{\mu \nu} = \del_{\mu} a_{\nu} - \del_{\nu} a_{\mu} + \theta^{\lam
 \rho} \del_{\lam} a_{\mu} \del_{\rho} a_{\nu} \ .
 \end{equation}
 Equations~\eqref{defsm} and~\eqref{deff} lead to   
 \begin{equation} \label{consf}
 F_{\mu \nu} (y(x)) = \left( {1 \ov 1 - f \theta } f \right)_{\mu \nu}
 (x) \ .
 \end{equation}
 Since 
 $$
 F_{\mu \nu} (k) = \int d^4 y \, F_{\mu \nu} (y) e^{i k \cdot y} \ ,
 $$ 
 we obtain 
 \begin{equation} \label{trans1}
 F_{\mu \nu} (k) = \int d^4 x \det M \left( {1 \ov 1 - f \theta} f
 \right)_{\mu \nu} e^{i k_{\rho} (x^{\rho} + \theta^{\rho \sigma}
 a_{\sigma})} 
 \end{equation}
where $M$ is the Jacobi matrix of the coordinate transformation 
from $y$ to $x$,
 i.e.
 \begin{equation}
 M_{\mu}^{\nu} = {\del y^{\nu} \ov \del x_{\mu} } 
 = \delta^{\nu}_{\mu} + \theta^{\nu \lam} \del_{\mu} a_{\lam} \ .
 \end{equation}
 Note that 
 \begin{equation} \label{jacobi}
 \theta^{\mu \nu} {\del y^{\lam} \ov \del x_{\mu} } 
 {\del y^{\rho} \ov \del x_{\nu} }  = \left[ \theta 
 (\theta^{-1}  - f ) \theta \right]^{\lam \rho} \;\;\;\; 
 \Rightarrow \;\;\;\;
 \det M = \sqrt{\det (1 - \theta f)} 
 \end{equation}
 From~\eqref{trans1} and~\eqref{jacobi} we finally get  
 \begin{equation} \label{transp}
 F_{\mu \nu} (k) = \int d^4 x \, \sqrt{\det (1 - \theta f)} 
 (\frac{1}{1 - f \theta} f )_{\mu \nu} \, 
 e^{i k_{\rho} (x^{\rho} + \theta^{\rho \sigma} a_{\sigma})} \ .
 \end{equation}
 
 In the above we have established an explicit  transformation between an 
 Abelian vector field $A$ with field strength $F$ and a ``Poisson'' 
 vector field $a$ whose field strength $f$~\eqref{pofs} has a nonlinear 
 term given by the Poisson bracket of $a$. It can be 
 checked~\cite{cornalba1,js,jsw} that an ordinary gauge transformation of 
 $A$ is equivalent to a gauge transformation of $a$ of the form
 \begin{equation} \label{lgt}
 a_{\mu} \rightarrow 
 a_{\mu} + \del_{\mu} \lam (x) - \{\lam, a_{\mu} \}_{B}
 = a_{\mu} + \del_{\mu} \lam (x) - \theta^{\nu \rho} \del_{\nu} \lam
 \del_{\rho} a_{\mu} \ . 
 \end{equation}
 Equations~\eqref{pofs} and~\eqref{lgt} are precisely 
 the formulas for the field strength and gauge 
 transformations of a noncommutative
gauge field $\hat{A}$ to first order in 
$\theta$ (see e.g. equations~\eqref{a6}). Thus~\eqref{transp} 
 gives us a closed form of the Seiberg-Witten map to first order in $\theta$.

 \subsection{A Proposal for the Seiberg-Witten Map} \label{sec:sw}
  
Now let us try to generalize~\eqref{transp} to all orders 
in $\theta$. First we replace $a$ in~\eqref{transp} by $\hat{A}$, 
and $f$~\eqref{pofs} by the noncommutative field strength 
$\hat{F}$,
\begin{equation} \label{defnf}
 \hat{F}_{\mu \nu}  =   \del_{\mu} \hat{A}_{\nu} - \del_{\nu}
 \hat{A}_{\mu} - i \hat{A}_{\mu} \ast \hat{A}_{\nu}
 + i \hat{A}_{\nu} \ast \hat{A}_{\mu}
\end{equation}
from which procedure we get 
\begin{equation} \label{pref}
 F_{\mu \nu} (k) = \int d^4 x  \sqrt{\det (1 - \theta \hat{F})} 
 \left( {1 \ov 1 - \hat{F}  \theta} \hat{F}
 \right)_{\mu \nu} e^{i k_{\rho} (x^{\rho} + \theta^{\rho \sigma}
 \hat{A}_{\sigma})} \ .
 \end{equation}
 Note that since $x$ satisfies the commutation relations 
 \begin{equation} \label{com}
 [x^{\mu},  x^{\nu}] = i \theta^{\mu \nu} 
 \end{equation}
 the exponential factor in equation~\eqref{pref} is nothing but a
 straight open Wilson line \cite{wadia}, i.e.
 \begin{equation} \label{matw}
  e^{i k_{\rho} (x^{\rho} + \theta^{\rho \sigma}
 \hat{A}_{\sigma})} = W(x,C) \ast e^{i k \cdot x}
 \end{equation}
where $C$ is given by~\eqref{path} and $W(x,C)$ by~\eqref{cwilson}. 
Substituting~\eqref{matw} into~\eqref{pref} we get 
\begin{equation} \label{prem}
F_{\mu \nu} (k) = \int d^4 x  \, \sqrt{\det (1 - \theta \hat{F})} 
\left( {1 \ov 1 - \hat{F}  \theta} \hat{F}
\right)_{\mu \nu} W(x, C) \, e^{i k \cdot x}
\end{equation}

The manipulations above are somewhat formal since we have not specified 
an ordering or product structure for the fields inside the integral.  
While equation~\eqref{prem} reduces correctly to~\eqref{transp} to first 
order in $\theta$, to find the exact form of the map, we need to
\begin{enumerate}
\item  specify an precise product structure and ordering 
for the integrand.

\item find any additional dependence on $\hat{F}$ or $\hat{A}$
which vanishes to first order in $\theta$.%
\footnote{More precisely, from~\eqref{transp}, 
by first order in $\theta$, we mean taking 
$\theta  \ll 1$ and $\theta \hat{A} \sim O(1)$.} 

\end{enumerate}

In~\cite{garousi,mehen}, the Seiberg-Witten map was found to cubic order 
in $\hat{A}$ while exact in $\theta$. Their results are
\begin{equation} \label{cubic}
\begin{split}
F_{\mu\nu} &= \hat{F}_{\mu \nu} +
\theta^{\lam \rho}\bigg( \partial_{\rho} 
(\hat{A}_{\lam} \ast^\prime \hat{F}_{\mu \nu }) + 
\frac{1}{2} \hat{F}_{\mu \nu}\ast^\prime \hat{F}_{\lam \rho} 
-\hat{F}_{\mu \lam}\ast^\prime \hat{F}_{\nu \rho} \bigg) \\ 
&+ \frac{1}{2}\theta^{\lam \rho} 
\theta^{\tau \sigma}\bigg(\partial_\lam \partial_\tau 
[\hat{F}_{\mu \nu} \, \hat{A}_\sigma \, \hat{A}_\rho]\ast_3 
- \partial_\tau [\hat{F}_{\lam \rho} \, 
\hat{F}_{\mu \nu} \, \hat{A}_\sigma]\ast_3 
+ 2 \partial_\tau [\hat{F}_{\mu \lam} \, \hat{F}_{\nu \rho} \, 
\hat{A}_\sigma]\ast_3 \bigg)\\
&-\theta^{\lam \rho}\theta^{\tau \sigma}\bigg(\frac{1}{2} 
[\hat{F}_{\mu \lam}\hat{F}_{\nu \rho}
\hat{F}_{\tau \sigma}]\ast_3- \frac{1}{8}[\hat{F}_{\mu \nu}\hat{F}_{\lam \rho}
\hat{F}_{\tau \sigma}]\ast_3- \frac{1}{4}[\hat{F}_{\mu \nu}\hat{F}_{\rho \tau}
\hat{F}_{\lam \sigma}]\ast_3 + 
[\hat{F}_{\lam \tau}\hat{F}_{\mu \sigma}\hat{F}_{\nu \rho}]\ast_3\bigg) \\
& + O(\hat{A}^4) \, .
\end{split}
\end{equation}

From our results in section~\ref{sec:owlast} it is easy to see 
that the gauge invariant completion of~\eqref{cubic}  is 
\begin{equation} \label{cubicc}
\begin{split}
F_{\mu\nu} (k)  &=  \int d^4 x  \,  e^{i k \cdot x} \, 
L_{\ast} \biggl [W(x, C) \, \biggl( \hat{F}_{\mu \nu} +
{1 \ov 2} \theta^{\lam \rho} \hat{F}_{\mu \nu} \hat{F}_{\lam \rho}
- \theta^{\lam \rho}\hat{F}_{\mu \lam}  \hat{F}_{\nu \rho} \\
& -\theta^{\lam \rho} \theta^{\tau \sigma}
\bigl( \frac{1}{2} \hat{F}_{\mu \lam} \hat{F}_{\nu \rho}
\hat{F}_{\tau \sigma} - \frac{1}{8} \hat{F}_{\mu \nu} \hat{F}_{\lam \rho}
\hat{F}_{\tau \sigma}- \frac{1}{4} \hat{F}_{\mu \nu}\hat{F}_{\rho \tau}
\hat{F}_{\lam \sigma} + 
\hat{F}_{\lam \tau}\hat{F}_{\mu \sigma}\hat{F}_{\nu \rho} \bigr) \biggr)
\biggr] \ ,
\end{split}
\end{equation}
where $L_{\ast}$ was defined in~\eqref{newg}. We further note 
that the $\hat{F}$ terms which multiply the open 
Wilson line in~\eqref{cubicc} are nothing but 
precisely the expansion of 
\begin{equation} \label{ins}
\sqrt{\det (1 - \theta 
 \hat{F})} \left( {1 \ov 1 - \hat{F}  \theta} \hat{F}
\right)_{\mu \nu}
\end{equation} 
to cubic order in $\hat{F}$. 

Thus by comparing~\eqref{prem} to~\eqref{cubicc} 
it is tempting to conjecture that the exact Seiberg-Witten map
is given by
\begin{equation} \label{ansatz}
F_{\mu \nu} (k) = \int d^4 x  \, L_{\ast} \left[ \sqrt{\det (1 - \theta 
\hat{F})} \left( {1 \ov 1 - \hat{F}  \theta} \hat{F}
\right)_{\mu \nu} W(x, C) \right] e^{i k \cdot x}
\end{equation} 
where the~\eqref{ins} part of the integrand should be understood as a 
power series of $\hat{F}$.  
Equation~\eqref{ansatz} is gauge invariant  by construction. It reduces 
to~\eqref{transp} when expanded to  first order in $\theta$ and 
when expanded to the third order in $\hat{A}$ while keeping the 
exact $\theta$-dependence it recovers~\eqref{cubic}.
We are currently unable to prove~\eqref{ansatz} to all orders in $\theta$.
However~\eqref{ansatz} does appear to be the simplest possibility 
that fits all the available information. We shall  see in the 
next section that~\eqref{ansatz} also gives rise to the correct formula 
when we ignore all the derivatives on $\hat{F}$.

One way to check that~\eqref{ansatz} is the right answer is to
write it in a form 
\begin{equation} \label{defca}
F_{\mu \nu} (k) = -i k_{\mu} A_{\nu}
+ i k_{\nu} A_{\mu} \ .
\end{equation} 
Since in the $U(1)$ case, $F_{\mu \nu}$  
is invariant under the noncommutative gauge transformations, $A_{\mu}$
found from~\eqref{defca} must  transform by an ordinary 
gauge transformation  under a noncommutative gauge transformation 
in $\hat{A}$.

Alternatively one may try to show that~\eqref{ansatz}
satisfies the set of differential equations derived in~\cite{sw} for 
$\hat{A}$ and $\hat{F}$ to ensure the gauge equivalence 
relations:
\begin{equation} \label{swe}
 \begin{split}
 \delta \hat{A}_{a} & = -{1 \ov 4} \delta \theta^{kl} 
 \left[ \hat{A}_k \ast (\del_l \hat{A}_a + \hat{F}_{la})
 + (\del_l \hat{A}_a + \hat{F}_{la}) \ast  \hat{A}_k 
 \right] \\
 \delta \hat{F}_{ab} & = {1 \ov 4} \delta \theta^{kl} 
 \left[ 2 \hat{F}_{ak} \ast \hat{F}_{bl} 
 + 2 \hat{F}_{bl} \ast \hat{F}_{ak} - \hat{A}_k \ast (\hat{D}_l \hat{F}_{ab}
 + \del_l \hat{F}_{ab}) - (\hat{D}_l \hat{F}_{ab}
 + \del_l \hat{F}_{ab}) \ast \hat{A}_k 
 \right]
 \end{split}
\end{equation}
However, the solutions of the equation~\eqref{swe} generally
depend on the choice of integration paths. Solutions corresponding to 
different paths are related by gauge transformations or field 
redefinitions (see e.g.~\cite{asakawa,tera}). Equivalently one may add 
terms which are gauge  transformations and field redefinitions to the 
right hand sides of~\eqref{swe}. That is, \eqref{swe} is just one 
of many possible ``Seiberg-Witten'' equations. 
Thus it is not very clear 
what the precise  prescription is for checking that~\eqref{ansatz}  
satisfies gauge equivalence.
As an example, it is easy to check that~\eqref{cubic} only 
satisfies~\eqref{swe} to quadratic order in $\hat{A}$ if we take the 
variation of $\theta$  to be proportional to itself, i.e. 
$\delta \theta^{ij} \propto \theta^{ij}$,%
\footnote{We note that to first 
order in $\theta$, it is indeed possible to check that~\eqref{transp}
satisfies the Seiberg-Witten equation~\eqref{swe} if we choose the path of
variation to be  $\delta \theta^{ij} \propto \theta^{ij}$.} 
while at cubic order 
one may have to use additional gauge transformations and field 
redefinitions\footnote{I would like to thank T. Mehen for correspondence 
about this point.}.

Since $\ast_n$ is fully symmetric with respect to
all its arguments, the prescription of~\eqref{ansatz} is reminiscent of the
symmetrized trace prescription of the non-Abelian Born-Infeld action
proposed by Tseytlin~\cite{tseytlin}. The trace here is over the
infinitely dimensional Hilbert space defined by~\eqref{com}. 
It would be interesting to investigate the
possible connection through the equivalence between the Born-Infeld
actions for the commutative and noncommutative gauge fields. We finally 
note that by using the definition of  $L_{\ast}$ and $\ast_{n}$ 
in the general $U(N)$ case, it is not inconceivable that if~\eqref{ansatz}
is correct, it may apply to the general $U(N)$ case.

 \section{Equivalence of Born-Infeld Actions Revisited} \label{sec:bi}
 
 In this section we give a direct proof of the equivalence  
 between the Born-Infeld actions for the ordinary and 
 noncommutative gauge fields, discovered in~\cite{sw}. We shall see
 that the explicit transformation~\eqref{ansatz} helps clarify 
 certain aspects of the equivalence.
 
 Since the Born-Infeld action describes the effective action of slowly
 varying fields on a D-brane, we can ignore terms involving
 derivatives of $\hat{F}$ in~\eqref{ansatz}. This means we can replace
 the $\ast_n$ products by ordinary products. However we
 shall still keep the exponential factor of $\hat{A}$ since 
 $\del_{\nu} \hat{A}_{\mu}$ is of the same order as 
 $\hat{F}$. Thus we now have
 \begin{equation} \label{preff}
 F_{\mu \nu} (k) = \int d^4 x \,  \sqrt{\det (1 - \theta \hat{F})} 
 \left( {1 \ov 1 - \hat{F}  \theta} \hat{F}
 \right)_{\mu \nu} e^{i k_{\rho} (x^{\rho} + \theta^{\rho \sigma}
 \hat{A}_{\sigma})}
 \end{equation}
 with ordinary products between various fields. Equation~\eqref{preff}
 is almost exactly the same as~\eqref{transp} except that here $\hat{F}$ 
 has the full nonlinear structure~\eqref{defnf}. 
 The steps from~\eqref{consf} to~\eqref{transp} suggests that we may
 derive an analogue of equation~\eqref{consf} for $\hat{F}$. 
 Define $X$ as 
 \begin{equation} \label{defx}
 X^{\mu} (x)  = x^{\mu} + \theta^{\mu \nu} \hat{A}_{\nu} (x) \ .
 \end{equation}  
 When $x$ satisfies the commutation relation~\eqref{com}, then
 \begin{equation} \label{comX}
 [X^{\mu}, X^{\nu}] = i \Theta^{\mu \nu} = 
 i \left( (1 - \theta \hat{F}) \theta \right)^{\mu \nu}
 \end{equation}
 In particular  
 \begin{equation} \label{meas} 
 \sqrt{\det \Theta} = \sqrt{\det \theta} \sqrt{\det (1 - \theta
 \hat{F})}
 \end{equation}
 and now~\eqref{preff} can be written as
 \begin{equation} \label{fouri} 
 F_{\mu \nu} (k) = \int \! d^4 x \,  {\sqrt{\det \Theta} \ov \sqrt{\det 
 \theta}}
 \left( {1 \ov 1 - \hat{F}  \theta} \hat{F}
 \right)_{\mu \nu} e^{i k \cdot X}
 \end{equation}
 
Recall that the integration over $x$ can be understood as taking the trace 
over the Hilbert space defined by~\eqref{com}, i.e.
 \begin{equation} \label{trace}
 \int d^4 x \, {1 \ov \sqrt{\det \theta}} \,\, \, \rightarrow 
 \,\,\, {\rm Tr}
 \end{equation} 
Considering~\eqref{comX} and~\eqref{trace} and ignoring any derivatives
on $\hat{F}$ we may rewrite~\eqref{fouri} as
\begin{equation} \label{bisw} 
 F_{\mu \nu} (k) = \int \! d^4 X \,  
 \left( {1 \ov 1 - \hat{F}  \theta} \hat{F}
 \right)_{\mu \nu} e^{i k \cdot X} \ .
 \end{equation}
This rather looks like a Fourier transformation in the variable $X$
and thus we find  
 \begin{equation} \label{tranw} 
 F_{\mu \nu} (X (x)) =   
 \left( {1 \ov 1 - \hat{F}  \theta} \hat{F}
 \right)_{\mu \nu} (x) 
 \end{equation}
Equation~\eqref{tranw} is our proposal for the Seiberg-Witten map 
in cases in which we can ignore the derivatives of $\hat{F}$ and  $F$.
In some sense~\eqref{tranw} is  very elegant in that it encodes the 
complicated relations between $F$ and $\hat{F}$ through a change of 
measure from $x$ to $X(x) = x + \theta \hat{A}$. 
In particular when $F = {\rm const}$ it reduces to the familiar 
result in~\cite{sw}.
 
In Appendix~\ref{sec:newproof} we present an alternative derivation of
equation~\eqref{tranw} based on the equivalence relation~\eqref{defD}
and the model of Catteneo-Felder~\cite{cf}. That the two  
approaches give the same result~\eqref{tranw} gives further  
support to our proposal~\eqref{ansatz}.

It is now a simple matter to demonstrate the equivalence of 
the Born-Infeld actions using~\eqref{tranw}. 
Using the relations between the open and closed string moduli~\cite{sw},   
\begin{equation} \label{moduli}
\begin{split}
&  {1 \ov g + B}  =  {1 \ov G + \Phi} + {\theta \ov 2 \pi \apr} \\
&  {1 \ov g_{s}} \sqrt{\det (g + B)}  =   {1 \ov G_s}  \sqrt{\det (G +
\Phi)} \ .
\end{split}
\end{equation}
we have  
\begin{equation} \label{bick}
\begin{split}
S_{BI} & = {1 \ov ( 2 \pi)^3 \apr^2 g_s} 
\int d^4 X \, \sqrt{\det (g + B + 2 \pi \apr F(X))} \\
& = {1 \ov ( 2 \pi)^3 \apr^2  g_s} 
\sqrt{\det (g + B)} \int d^4 x \,  \sqrt{\det (1 - \theta \hat{F})} 
\sqrt{\det \left( 1 + {1 \ov g + B} 2 \pi \apr {1 \ov 1 -
\hat{F} \theta} \hat{F} \right)} \\
& = {1 \ov ( 2 \pi)^3 \apr^2 G_s}  \sqrt{\det (G + \Phi)} \\
& \times \int d^4 x \,  
\sqrt{\det (1 - \theta \hat{F})} \, 
\sqrt{\det \left( 1 + ({1 \ov G + \Phi} + {\theta \ov 2
\pi \apr}) 2 \pi \apr \hat{F} {1 \ov 1 -
\theta \hat{F}}  \right)} \\
& =  {1 \ov ( 2 \pi)^3 \apr^2 G_s} \int d^4 x \, 
\sqrt{\det (G + \Phi  + 2 \pi \apr \hat{F})} 
\end{split}
\end{equation}
In the second line above we have used  
equation~\eqref{tranw} and made a coordinate change inside the integral. 
In the third line we substituted the relations~\eqref{moduli}.

In~\cite{sw}, the equivalence between the ordinary and noncommutative 
Born-Infeld Lagragians was proved by showing that the noncommutative
BI Lagrangian is independent of the $\theta$ parameter up to total 
derivative terms. It was pointed out there that, for constant $F$, where 
the map is given by
\begin{equation} \label{constf}
 F_{\mu \nu}  =   
 \left( {1 \ov 1 - \hat{F}  \theta} \hat{F}
 \right)_{\mu \nu} 
\end{equation}
the two Lagrangians are actually different. The reason is that for 
constant $F$ it is not legitimate to throw away total derivatives 
which may contribute through boundary terms\footnote{The discussion of 
this paragraph is developed with S. Minwalla.}. 
What~\eqref{bick} says is that the information lost in the boundary 
terms is precisely encoded in the measure change between $X$ and $x$, 
and when included correctly, the two actions are indeed equivalent.     
Consider a simple example with $\hat{F} = {1 \ov \theta}$
in a finite region of space and zero outside.
From equation~\eqref{constf} the ordinary $F$ field is infinite in this limit
and the two BI lagrangians are obviously different. However the measure
change from $x$ to $X$ indicates the region in which $F$ is nonzero shrinks to
zero size at the same time, ensuring that the whole actions are  equivalent.

\section{Discussion} \label{sec:conc}
 
In this paper we have discussed that the natural way 
to understand the appearance of the $\ast_n$ operations 
in various places of  noncommutative gauge theories 
is through the open Wilson lines. These include 
the one-loop effective action of noncommutative gauge theories, 
the couplings between massless closed and open string modes, and 
the Seiberg-Witten map between the ordinary and noncommutative 
Yang-Mills fields.  One common theme in the discussion is the 
need for gauge invariance and the Wilson line is the natural object
to realize that. In all cases the $L_{\ast}$ 
prescription~\eqref{newg}---
i.e. smearing operators along a straight open Wilson line---played an  
important role. In one-loop amplitudes and closed-open string couplings 
the $L_{\ast}$ prescription can be understood from the ``stretched string
effect'' discussed in~\cite{lm,lm2} and has its origin in the 
integrations over the vertex operator insertions on the worldsheet.
It should be interesting to understand the physical reason for its appearance
in the Seiberg-Witten map.
Of course, the above cases  are not unrelated 
to each other, e.g. one-loop open string amplitudes are related  
to the closed-open tree-level amplitudes by factorization. 
Also since closed string modes have a simple off-shell
coupling to ordinary Yang-Mills field variables (e.g. through ordinary 
Born-Infeld action), it is not hard to imagine that the presence 
of open Wilson lines in the closed-noncommutative Yang-Mills field 
couplings and the Seiberg-Witten map may be of the same origin.

The explicit couplings worked out in section~\ref{sec:f4} between 
the modes of noncommutative gauge theory and massless 
closed strings in flat space should be useful for understanding 
the operator-field correspondence in the supergravity description
of the theory. Based on the examples found in this paper,
we may try to speculate what is the general pattern for the 
operator-field matching.
A most naive expectation would be that we  start with 
operator-field matching in the ordinary AdS/CFT correspondence, replace the
Yang-Mills field variables by their noncommutative counterparts 
and attach the resulting operators to a straight Wilson line with 
$L_{\ast}$ ordering to obtain the field-operator matching in the 
noncommutative case. It should be interesting to  work out more 
examples by using supersymmetry or by 
considering amplitudes involving the scalar fields and 
fermions.
However we caution that the situation might be much more complicated
since in the noncommutative case the supergravity background~\cite{hi,mr} 
has rather complicated mixings between small fluctuations due to non-trivial 
background fields. This might imply that the operator-field matching 
in the supergravity description  is more intricate than the couplings we 
observed between the noncommutative SYM and closed string modes 
in {\em flat} space.

The closed form of the Seiberg-Witten map and its cousin in the
slowly-varying field case can also have various applications.
For example it should be useful for studying questions like 
the behaviors of solitons in the presence of a 
$B$-field (e.g.~\cite{sw,mmms,tera2}) 
and constraining the structure of higher-derivative terms in the 
Born-Infeld action (e.g.~\cite{okawa,cornalba2}).
From the string theory point of view, the map follows from that ordinary and 
noncommutative field variables arise from different regularizations of 
the same string worldsheet theory and their respective 
effective actions correspond to different off-shell extensions 
of the on-shell string amplitudes. An explicit form of the map 
in the field theory context might be helpful for studying the more difficult 
question of  finding the relations between different off-shell extensions 
in  string theory, a question which is of much importance in string field 
theory. 

\acknowledgments
 
I have benefited from useful discussions and correspondences with
T. Banks, M.~Douglas, R.~Gopakumar, D.~Gross, C.~Hofman, G.~Moore, 
M.~Mari\~no, T.~Mehen, J. Michelson, S.~Minwalla, B.~Pioline, S.-J.~Rey, 
M.~Rozali and P.~Schupp. I thank J.~Michelson for encouragement and for 
a critical reading  of the manuscript. The 
author also thanks the Harvard group, where a portion of this 
work was performed, for hospitality and fruitful discussion.  
This work was supported by DOE grant
\hbox{\#DE-FG02-96ER40559}.

\appendix

\section*{Appendix}
 
\section{Notations and Conventions} \label{sec:notation}

Here we list our conventions about noncommutative gauge theory.
We will restrict our discussion to a noncommutative ${\mathbb{R}}^4$,
in which 
\begin{eqnarray}\label{comm}
[ x^\mu, x^\nu ] = i \, \theta^{\mu \nu} \,.
\end{eqnarray}
where c-number $\theta$ is antisymmetric and non-degenerate.
The algebra of functions on this space is given by 
the $\ast$-product:
\begin{eqnarray}\label{star}
f(x) \ast g(x) = \exp \left[\frac{i}{2} \, \theta^{\mu \nu} 
{\partial \over \partial x^\mu} 
{\partial \over \partial y^\nu} \right] f(x) g(y) \bigg \bracevert_{x=y} \, .
\end{eqnarray}
In momentum space,
\begin{equation} \label{astm}
f(k_1) \ast g(k_2) = 
f(k_1) g (k_2) \exp 
\left (- \frac{i}{2} k_1 \times k_2 \right) 
\end{equation}
where 
\begin{equation}
k_1 \times k_2 \equiv k_{1 \mu} \theta^{\mu \nu} k_{2 \nu}
\end{equation}
and the right hand side of the equation~\eqref{astm} is given by the 
ordinary product.

The action for a noncommutative gauge theory  is
\begin{eqnarray}\label{action}
S = -{1 \over 4 g^2} \int d^4x \, {\rm Tr} \hat{F}_{\mu \nu} 
\ast \hat{F}^{\mu \nu} \, ,
\end{eqnarray}
where the noncommutative gauge field strength is 
\begin{eqnarray}\label{fieldstrength}
\hat{F}_{\mu  \nu} = \partial_\mu \hat{A}_\nu- \partial_\nu \hat{A}_\mu 
- i \hat{A}_{\mu} \ast \hat{A}_\nu + i \hat{A}_{\nu} \ast \hat{A}_\mu \ .
\end{eqnarray}
Under an infinitesimal  gauge transformation,
\begin{eqnarray}\label{tran}
\delta_{\hat{\lambda}} \hat{A}_\mu & = & \partial_\mu \hat{\lambda} + 
i \hat{\lambda} \ast \hat{A}_\mu - i \hat{A}_\mu \ast \hat{\lambda} \, , \\
\delta_{\hat{\lambda}} \hat{F}_{\mu \nu} & = &
i \hat{\lambda} \ast \hat{F}_{\mu \nu} - i  \hat{F}_{\mu \nu} \ast 
\hat{\lambda}
\, . \,\,\,\,\,\,\, \nonumber
\end{eqnarray}
 
To first order in $\theta$ the above formulas for field strength and 
gauge transformations become
\begin{equation} \label{a6}
\begin{split}
\hat{F}_{\mu  \nu} & = \partial_\mu \hat{A}_\nu- \partial_\nu \hat{A}_\mu
+  \theta^{\lam \rho} \del_{\lam} \hat{A}_{\mu}  \del_{\rho} \hat{A}_{\nu} 
= \partial_\mu \hat{A}_\nu- \partial_\nu \hat{A}_\mu + \{\hat{A}_{\mu},
\hat{A}_{\nu} \}_{\theta^{-1}} \\
\delta_{\hat{\lambda}} \hat{A}_\mu & =  \partial_\mu \hat{\lambda} -
\theta^{\rho \sigma} \del_{\rho} \hat{\lam}  \del_{\sigma} \hat{A}_{\mu} 
=  \partial_\mu \hat{\lambda} - \{\hat{\lam},\hat{A}_{\mu} \}_{\theta^{-1}}
\\
\delta_{\hat{\lambda}} \hat{F}_{\mu \nu} & = - 
\theta^{\rho \sigma} \del_{\rho} \hat{\lam}  
\del_{\sigma} \hat{F}_{\mu \nu} 
= -\{\hat{\lam}, \hat{F}_{\mu \nu}\}_{\theta^{-1}} \ . 
\end{split}
\end{equation}
where $\{\, \}_{\theta^{-1}}$ denotes the Poisson bracket with respect to
the symplectic form $(\theta^{-1})_{\mu \nu}$.

A Wilson line is given by
\begin{smaleq}
\begin{equation} \label{dwilson}
\begin{split}
& W(x,C) \\
& =  P_{\ast} \exp \left( i 
\int_0^1 d \sigma \partial_{\sigma} \,
\xi^{\mu} (\sigma) \, \hat{A}_{\mu} (x + \xi (\sigma))
\right)  \\
& =  \lim_{\Delta x_j \rightarrow 0} \prod_{j} \,  [1 + i \hat{A} ({x_j})
\cdot \Delta x_j]_{\ast} \\ 
& =  
\sum_{n=0}^{\infty} i^n \int_0^1 \! d \sigma_1 \int_{\sigma_1}^1
\! d \sigma_2 \cdots  \int_{\sigma_{n-1}}^1
\! d \sigma_n \, \del_{\sigma_1} \xi^{\mu_1} \, \cdots 
\del_{\sigma_n}\xi^{\mu_n} \,
\hat{A}_{\mu_1}(x + \xi(\sigma_1)) \ast
\cdots \ast \hat{A}_{\mu_n}(x + \xi(\sigma_n)) 
\end{split}
\end{equation}
\end{smaleq}

\section{The $\ast_n$ Operations} \label{sec:exam}

$\ast_n$ is an $n$-ary operation defined on the space 
of $n$-functions~\cite{lm2}. Its introduction was motivated from 
the structure of the one-loop non-planar amplitudes of the noncommutative 
gauge theories; for more details see~\cite{lm2,lm3}. 

It is convenient to define the $\ast_n$ operation  in the 
momentum space, i.e. 
\begin{equation} 
\ast_n [f_1 (x), \cdots, f_n (x)]  = \prod_{i=1}^n 
\left(\int {d ^4 k_i \ov (2 \pi)^4} \right) \,
e^{i (k_1 + \cdots + k_n) \cdot x} \, 
\ast_n \left[ f_1(k_1), f_2(k_2), \cdots, f_n (k_n) \right] \ .
\end{equation}
In the $U(1)$ case, we define:
\begin{equation} 
\ast_n \left[ f_1(k_1), f_2(k_2), \cdots, f_n (k_n) \right] 
 = f_1(k_1) \, f_2(k_2) \, \cdots \, f_n (k_n) \, J_{n} (k_1, \cdots, k_n)
\end{equation} 
where the right hand side of the equation  
is given by the ordinary product and $J_n$ is 
($\tau_{ij} = \tau_i - \tau_j$):
\begin{equation} \label{amjn}
J_n (k_1, \cdots, k_n) = \int_{0}^{1} d \tau_1 \cdots 
\int_{0}^{1} d \tau_n \; \exp \left[ - \frac{i}{2} \sum_{i<j}^n
(k_i \times k_j) (2 \tau_{ij} - \epsilon (\tau_{ij})) \right] 
\end{equation}
with $k_1 \times k_2 \equiv k_{1 \mu} \theta^{\mu \nu} k_{2 \nu}$.
In~\eqref{amjn}, the integrations over $\tau$ have their origin, in  
string theory amplitudes, as integrations over the vertex operator 
insertions along the worldsheet boundary. The integrand of~\eqref{amjn} comes 
from the $\theta$-dependent part of the annulus propagators.
To compare with the Moyal product, we note that in the momentum space 
the Moyal product of $n$ functions is given by
\begin{equation} 
f_1 (k_1) \ast f_2 (k_2) \ast \cdots \ast f_n (k_n)
=  f_1(k_1) \, f_2(k_2) \, \cdots \, f_n (k_n) \, 
\exp \left[ - \frac{i}{2} \sum_{i<j}^n
(k_i \times k_j) \right]
\end{equation}

In the general $U(N)$ case, where $f_i$ are $N \times N$-valued matrices,  
let $f_i  = \sum_{a_i} f_i^{a_i} T^{a_i}$ where $T^{a_i}$ are a set of basis.
Then 
\begin{equation}
\ast_n \left[ f_1(k_1), f_2(k_2), \cdots, f_n (k_n) \right] 
= f_1^{a_1}(k_1)\,  f_2^{a_2}(k_2)\, \cdots \, f_n^{a_n} (k_n) \,  
J_{n} (a_1,k_1; \cdots; a_n,k_n)
\end{equation} 
with
\begin{equation} \label{defnaj}
\begin{split}
& J_n (a_1,k_1; \cdots; a_n, k_n)  \\
& = \left( \prod_{i=1}^n \int_{0}^{1} d \tau_i \right) \,
P_{\tau} \left( T^{a_1} \cdots T^{a_n} \right) \,
\exp \left[ - \frac{i}{2} \sum_{i<j}^n
(k_i \times k_j) (2 \tau_{ij} - \epsilon (\tau_{ij})) \right]
\end{split}
\end{equation}
where $P_{\tau}$ denotes an ordering of matrices $T^{a_i}$ according to the 
ordering of $\tau_i$ and is motivated from the structure of the
string amplitudes with Chan-Paton factors. 

Here we list the explicit expressions
for $n=2,3$ in the $U(1)$ case which were found in \cite{garousi,lm}. 
When $n=2$, 
\begin{equation}
J_2 = { \sin{k_1 \times k_2 \ov 2} \ov {k_1 \times k_2 \ov 2 }},
\;\;\;\;\;\;
\ast_2 (f(x), g(x)) \equiv f(x) 
\frac{\sin\bigl(\frac{1}{2}
\theta^{\mu\nu}\lvec[\mu]{\p}\rvec[\nu]{\p}\bigr)}%
{\frac{1}{2}\theta^{\mu\nu}\lvec[\mu]{\p}\rvec[\nu]{\p}}
g(x) 
\end{equation}
When $n=3$,
\begin{equation} \label{u1totnonplanar31}
J_3 (k_1, k_2, k_3) = \frac{\sin\bigl(\frac{k_2\times k_3}{2}\bigr) 
\sin\bigl(\frac{k_1\times (k_2+ k_3)}{2}\bigr)}%
{\frac{(k_1+k_2)\times k_3}{2} \frac{k_1\times(k_2+k_3)}{2}}
+\frac{\sin\bigl(\frac{k_1\times k_3}{2}\bigr) 
\sin\bigl(\frac{k_2\times (k_1+ k_3)}{2}\bigr)}%
{\frac{(k_1+k_2)\times k_3}{2} \frac{k_2\times(k_1+k_3)}{2}}.
\end{equation}
and the corresponding $\ast_3$ is 

\begin{multline} \label{ternary}
\ast_3[f(x),g(x),h(x)] \\* \equiv 
\left[
\frac{\sin\bigl(\frac{\partial_2\times \partial_3}{2}\bigr) 
\sin\bigl(\frac{\partial_1\times (\partial_2+ \partial_3)}{2}\bigr)}%
{\frac{(\partial_1+\partial_2)\times \partial_3}{2} 
\frac{\partial_1\times(\partial_2+\partial_3)}{2}}
+ \frac{\sin\bigl(\frac{\partial_1\times \partial_3}{2}\bigr) 
\sin\bigl(\frac{\partial_2\times (\partial_1+ \partial_3)}{2}\bigr)}%
{\frac{(\partial_1+\partial_2)\times \partial_3}{2}
\frac{\partial_2\times(\partial_1+\partial_3)}{2}}\right]
\evalat{f(x_1) g(x_2) h(x_3)}{x_i=x},
\end{multline}

\section{An Alternative Derivation of the Seiberg-Witten Map 
in the Slowly-varying Field Case} \label{sec:newproof}

Here we give a different derivation of~\eqref{tranw} of section~\ref{sec:bi}
based on the equivalence relation~\eqref{defD} and the path integral 
representation of Kontsevich's star product derived in~\cite{cf}. The 
derivation is very similar to that used in section~\ref{sec:poisson}.

It was shown in~\cite{cf} that when a  symplectic form $\omega$ 
is non-singular, the Kontsevich's star-product can be 
obtained from the following path integral\footnote{What we consider 
below is a special simplified case of more general discussions in~\cite{cf}.},
\begin{equation} \label{gstar}
f(y) \ast_{\omega} g(y) = \int_{Y(\infty) = y} 
DY \, f (Y(1)) g(Y(0)) \, e^{i S} 
\end{equation}
where $Y$ defines  a two-dimensional field theory on a disk
and $0,1,\infty$ are three distinct ordered 
points on the boundary of the disk. In~\eqref{gstar} the action is given by
\begin{equation} \label{cfaction}
S = {1 \ov 2} \int d^2 \sigma \, \epsilon^{ab} \omega_{\mu \nu} 
\del_a Y^{\mu} \del_{b} Y^{\nu} 
\end{equation}

In our case, from~\eqref{defsm}, 
$$
\omega = (\theta^{-1})_{\mu \nu} + F_{\mu \nu} 
$$ 
is an exact form  and the action~\eqref{cfaction} reduces 
to an action defined on the boundary of the disk,
\begin{equation} \label{baction}
S = {1 \ov 2} \int d \sigma \,  
\left[ (\theta^{-1})_{\mu \nu} 
Y^{\mu} \del_{\sigma} Y^{\nu} + A_{\mu} (Y(\sigma)) \del_{\sigma} Y^{\nu} 
\right]
\end{equation} 
where $A_{\mu}$ is the gauge field for $F$ 
and~\eqref{baction} can be considered as the action of a 
string world-sheet action in the presence of a gauge field background in the 
Seiberg-Witten decoupling limit. 

Let $Y^{\mu} (\sigma) = y^{\mu}  + \xi^{\mu} (\sigma)$.   
When we can ignore the derivatives on $F$, the action can be simplified 
to yield 
\begin{equation} \label{sbaction}
S = {1 \ov 2} \int d \sigma 
\left[\theta^{-1}_{\mu \nu} + F_{\mu \nu} (y) \right] \,  
\xi^{\mu} \del_{\sigma} \xi^{\nu} + O(\del F)
\end{equation}
In this limit the structure of the star-product is also considerably 
simplified. Let us take  $f = y^{\mu}$ and $g= y^{\nu}$ in~\eqref{gstar},
which leads to
\begin{equation}
y^{\mu} \ast_{\omega} y^{\nu} = {i \ov 2} (\omega^{-1})^{\mu \nu}
\end{equation}
With equation~\eqref{defD} this implies 
\begin{equation}
{\cal D} \left( [y^{\mu}, y^{\nu}]_{\ast_{\omega}} \right)
= [{\cal D} y^{\mu} , {\cal D} y^{\nu}]_{\ast_B }
\end{equation}
Now using~\eqref{defA}, we find that
\begin{equation}  
 i (\omega^{-1})_{\mu \nu} (X (x)) =   [X^{\mu}, X^{\nu}]_{\ast}
\end{equation}
and
\begin{equation}  
 F_{\mu \nu} (X (x)) =   
\left( {1 \ov 1 - \hat{F}  \theta} \hat{F}
 \right)_{\mu \nu} (x) 
\end{equation}

 \end{document}